\documentclass{aastex6}

\usepackage{multirow}
\fullcollaborationName{The Friends of AASTeX Collaboration}

\begin{document}
\title{Detection of GeV $\gamma$-Ray Emission from supernova remnant SNR G15.9+0.2 with Fermi-LAT}

\author{Yunchuan Xiang\altaffilmark{1}, Zejun jiang\footnote{Corresponding author}\altaffilmark{1} and Mengyao Tang\altaffilmark{1}}

\altaffiltext{1}{Department of Astronomy, Yunnan University, and Key Laboratory of Astroparticle Physics of Yunnan Province, Kunming, 650091, China, xiang{\_}yunchuan@yeah.net, zjjiang@ynu.edu.cn}

\begin{abstract}
We first report GeV $\gamma$-ray emission from supernova remnant (SNR) G15.9+0.2 in this work. The results show  that its power-law spectral index is  2.94$\pm$0.25 with a 6.47$\sigma$ significance level, and the $\gamma$-ray emission can be characterized by a two-dimensional (2D) Gaussian spatial distribution, which has a better improvement than the case of a point source. Moreover, we find that its likely counterparts from the radio, X-ray, and TeV  energy bands are well coincident with its spatial location. We suggest that the new $\gamma$-ray emission may originate from SNR G15.9+0.2. Analyzing the variability from 12.4 years of the light curve (LC), we identify  that this LC exists weak variability with a 2.69$\sigma$ variability significance level. We investigated the 2D Gaussian extended region and did not identify certified  active galactic nuclei from the region of this SNR; thus, we suggest that the new $\gamma$-ray emission may originate from SNR G15.9+0.2. On this basis, we discussed the probable origins of its $\gamma$-ray radiation from leptonic and hadronic scenarios, respectively.
\end{abstract}

\section{Introduction} \label{sec:intro}
Supernova remnant (SNR) is considered to be an efficient cosmic-ray factory. After the explosion of SNR, it is considered that the 10\% kinetic energy of SNR transferred to CRs, and the maximum energy of cosmic-ray particles can be accelerated to approximately 10$^{15}$ eV through the diffusive shock acceleration mechanism \citep{Bell1978,Blandford1987,Drury1994, Morlino2012}. 
Their multiband spectra exhibit a typical bimodal structure through leptonic or/and hadronic processes \citep[e.g., ][]{Zeng2019,Zeng2021}. 
For the radio-to-X-ray band, it is generally recognized that synchrotron radiation dominates \citep{Allen1997,Allen1999,Uchiyama2007}. 
For the GeV-to-TeV energy band, inverse Compton scattering and bremsstrahlung of relativistic electrons are generally considered to be important mechanisms \citep{Vink2012}.  In addition, decay of neutral pions  produced in the inelastic hadronic interaction becomes more and more important to explain the GeV and TeV emissions of SNRs \citep[e.g.,][]{Xin2017,Xin2019,Yang2021,Xiang2021}. 
Detection of GeV $\gamma$-ray emission of SNR is very important for evaluating SNR's contribution to cosmic-ray flux in Milky Way \citep{Acero2016}, and it can also help us explore the acceleration mechanism of cosmic-ray particles and limit the energy distribution of accelerated particles, which provides further understanding of the evolution process of cosmic-ray particles in SNRs  \citep{Zhang2007,Finke2012,Tang2013}. 
Thus far, only 24 SNRs have been firmly certified in the Fermi Large Telescope Fourth Source Catalog \citep[4FGL;][]{Abdollahi2020}.  Therefore, more GeV SNRs are required to recognize the nature of particle acceleration within SNRs.


\citet{Caswell1982} found a clear shell structure from SNR G15.9+0.2 at 1415 MHz with about 58$^{''}$ resolution using the Fleurs synthesis radio telescope. 
Based on the observations at 327.5 MHz and 1425 MHz from the NRAO VLA sky survey (NVSS), \citet{Dubner1996} found that the  eastern border of SNR G15.9+0.2 had a bright shell feature; the northwest edge had two fainter knots; the north of the shell appeared an extension region of the radio emission with a 3$\sigma$ noise level at 1425 MHz.
Through XMM-Newton observations, \citet{Maggi2017} found the evidence of spatial variations by measuring the Fe K line features. They believed that SNR G15.9+0.2 with the lowest Fe K centroid energy is the core-collapse SNR. Moreover, they identified ejecta emission from this SNR by observing the Fe K line features. They identified that the progenitors of SNR G15.9+0.2 originated from a massive star, according to the observation the abundance ratios of Ca, Ar, S, Si, and Mg. 

For the GeV band data, \citet{Acero2016} did not find significant GeV $\gamma$-ray emission of the SNR in the 1-100 GeV energy band using the Fermi Large Area Telescope (Fermi-LAT), and they provided its upper limits with the 95\% and 99\% confidence level for the power-law spectral indices of 2.0 and 2.5, respectively. 
For the TeV energy band, 
 \citet{Abeysekara2017} found that the TeV source 2HWC J1819-150 is closer to
  SNR G15.9+0.2 with a position separation of 0$^{\circ}$.1.  
In addition, the differential flux at 7 TeV is 59.0$\pm$7.9$\times$10$^{-15}$ TeV$^{-1}$ cm$^{-2}$ s$^{-1}$ with a spectral index of -2.88$\pm$0.10.

In this study, for the improvement and accumulation from the Fermi-LAT Pass 8 data and the update of $\gamma$-ray background models from galactic diffuse emission and the isotropic extragalactic emission \citep{Abdollahi2020}, the GeV $\gamma$-ray emission of SNR G15.9+0.2 was reanalyzed using approximately 12.4 years of the Pass 8 data. In the preliminary analysis, a likely GeV $\gamma$-ray emission from the region of SNR G15.9+0.2 was found, which strongly inspired us to explore the characteristics and origin of this GeV $\gamma$-ray emission in this study. Subsequent works include the introduction of data reduction in Section 2, the presentation of the analysis results in Section 3, and the discussion and conclusion about the likely origins of the GeV radiation in Section 4.

\section{Data Reduction}

Using Fermitools version {\tt v11r5p3}\footnote{http://fermi.gsfc.nasa.gov/ssc/data/analysis/software/}, we analyzed the GeV $\gamma$-ray emission from the region of SNR G15.9+0.2 by selecting the instrumental response function (IRF) ``P8R3{\_}SOURCE{\_}V3'' and the Pass 8 ``Source'' event class (evtype = 3 and evclass = 128).  The observation period was selected to be from August 4, 2008, to December 29, 2020 (mission elapsed time (MET) 239557427-630970757). The energy range was selected to be 1-500 GeV to reduce  contamination from the galactic diffuse emission for a large point spread function in the low-energy band.
Photon events with the maximum zenith angles of $90^{\circ}$ were selected to suppress the pollution from the Earth Limb. 
A $20^{\circ}\times 20^{\circ}$ region of interest (ROI), centered at the position from SIMBAD (R.A., decl.= 274$^{\circ}$.72, -15$^{\circ}$.03) \footnote{http://simbad.u-strasbg.fr/simbad/}, was selected for this analysis.
We selected the script {\tt make4FGLxml.py}\footnote{https://fermi.gsfc.nasa.gov/ssc/data/analysis/user/} to generate a source model file, and sources from the 4FGL within the ROI of 30$^{\circ}$ were included in the model file.
Then, we included a point source with a power-law spectrum at the SIMBAD location of SNR G15.9+0.2 to the model file.
The binned likelihood tutorial\footnote{https://fermi.gsfc.nasa.gov/ssc/data/analysis/scitools/binned{\_}likelihood{\_}tutorial.html} was followed in the analysis.
Furthermore, spectral indexes and normalizations from sources within the $5^{\circ}$ range of the  ROI were set as free in the model file. The normalizations from the isotropic extragalactic emission ({\tt iso{\_}P8R3{\_}SOURCE{\_}V3{\_}v1.txt}) and the galactic diffuse emission ({\tt gll{\_}iem{\_}v07.fits})\footnote{http://fermi.gsfc.nasa.gov/ssc/data/access/lat/BackgroundModels.html} were also set as free.

\section{Source Detection} \label{sec:data}

\begin{figure}
\begin{center}
\includegraphics[scale=0.40,angle=0]{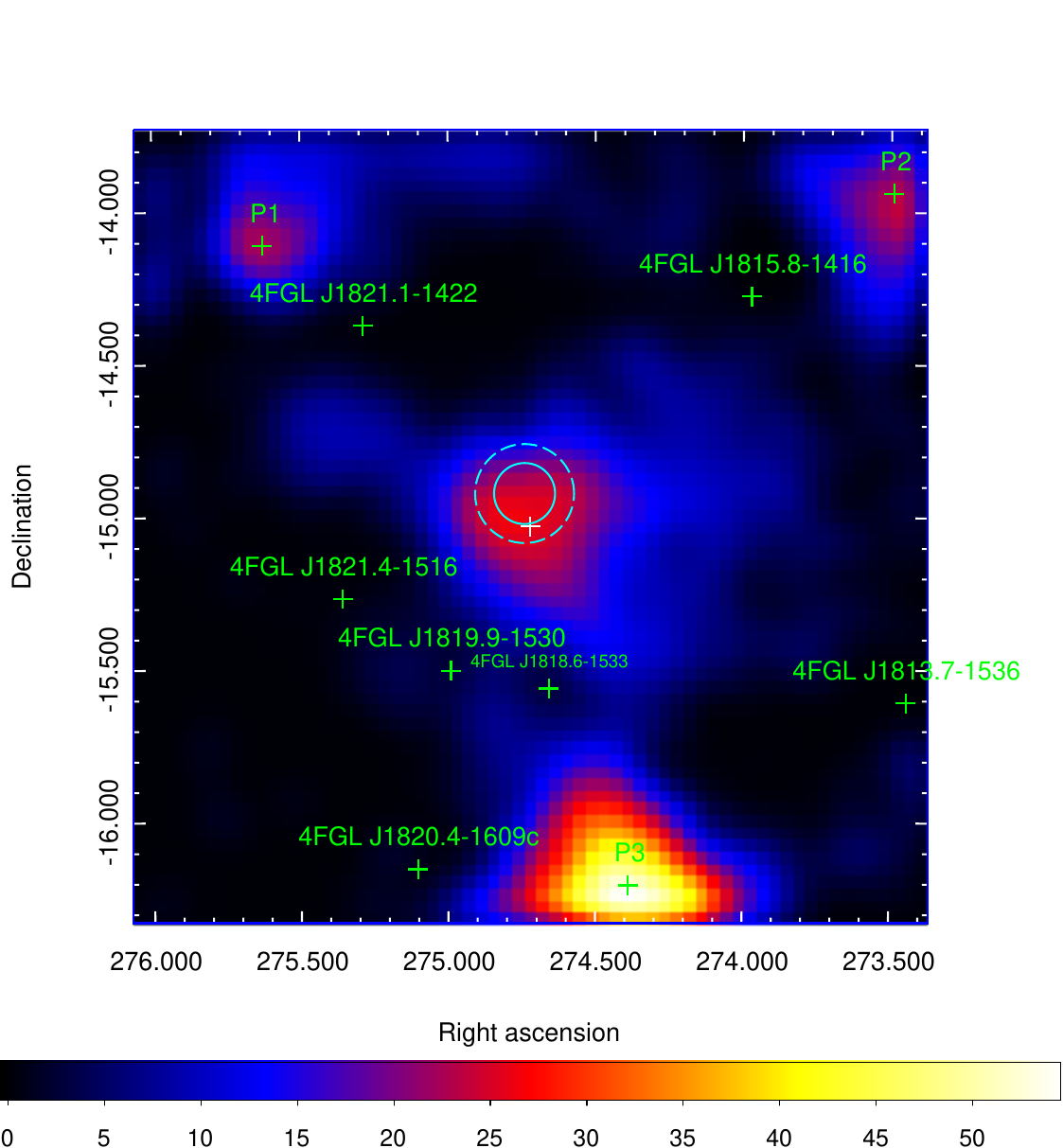}
\includegraphics[scale=0.40,angle=0]{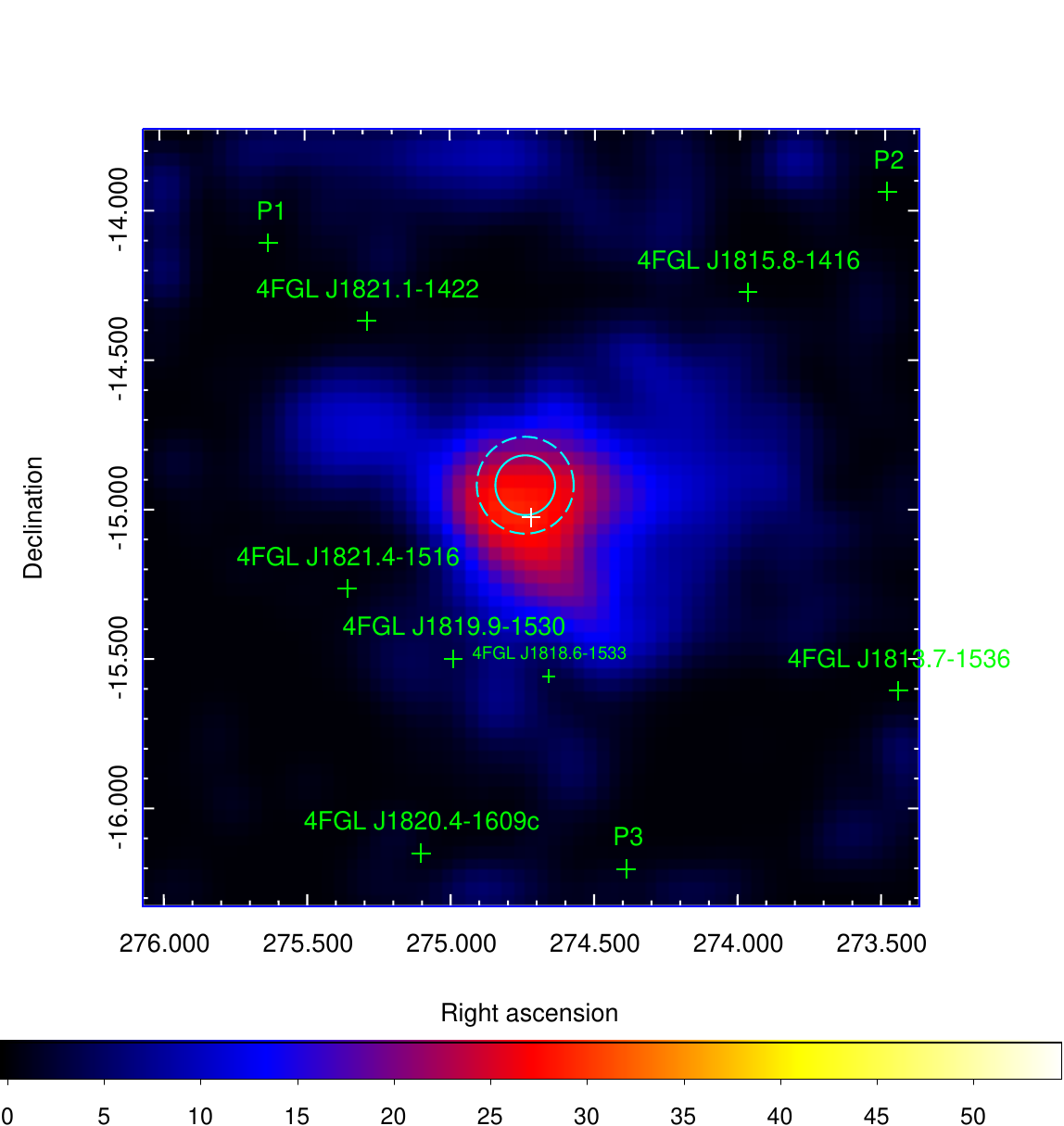}
 \caption{
These TS maps of $2^{\circ}.6\times2^{\circ}.6$, which were smoothed with a Gaussian function with a kernel radius of $0^{\circ}.3$, in the 1-500 GeV energy band with a 0$^{\circ}$.04 pixel size centered at the SIMBAD position of SNR G15.9+0.2 marked as a white cross. 
The 68\% and 95\% error circles of the best-fit position of this 
SNR are marked by using solid and dashed cyan circles in these TS maps,  respectively.
Left panel: TS map including all the 4FGL sources and the residual radiations in the region.
Right panel: TS map after deducting three $\gamma$-ray excesses including P1-P3.}
\label{Fig1}
\end{center}
\end{figure}

Running the command {\tt gttsmap}, the test statistic (TS) map, which is centered at the SIMBAD location of SNR G15.9+0.2 in the 1-500 GeV energy band, was first calculated in this analysis. In the left panel of Figure \ref{Fig1},  
 significant $\gamma$-ray radiation with a TS  value = 30.20 was found in the region of SNR G15.9+0.2. 
Here, the TS value, defined as TS = 2log(L$_{1}$/L$_{0}$) from \citep{Mattox1996}, is calculated to quantify a significant source, and L$_{1}$ and L$_{0}$ represent maximum-likelihood values; L$_{1}$ contains target source; L$_{0}$ does not contain.
In addition, we identified three significant $\gamma$-ray excesses from the locations of P1, P2, and P3. Then, we chose to add these three point sources, with power-law spectra in the local maxima of the TS map, to their locations\footnote{The location of P1: (R.A., decl.=275$^{\circ}$.63, -14$^{\circ}$.11); that of P2: (R.A., decl.=273$^{\circ}$.49, -13$^{\circ}$.99); that of P3: (R.A., decl.=274$^{\circ}$.39, -16$^{\circ}$.21).} to subtract the three significant residual emissions within the $2^{\circ}.6 \times 2^{\circ}.6$ TS map for all subsequent analyses. As shown in the right panel of Figure \ref{Fig1}, the $\gamma$-ray radiation was still  significant with the TS value of $29.61$ in the region of SNR G15.9+0.2.

\begin{figure}
\centering
  \includegraphics[scale=0.41,angle=0]{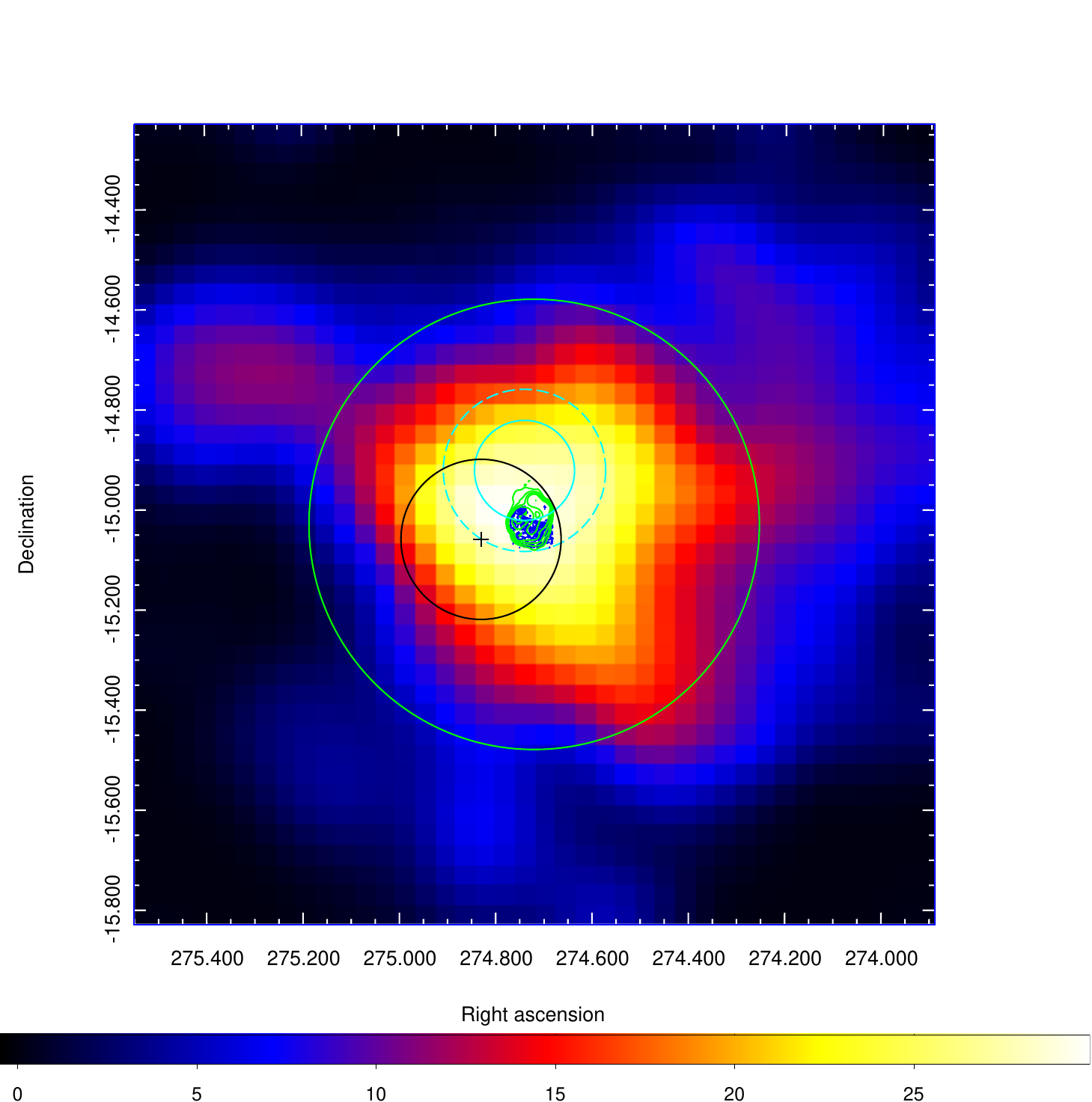}
 \caption{TS map of the region of 1$^{\circ}$.6 $\times$ 1$^{\circ}$.6 is smoothed with a Gaussian kernel of 0$^{\circ}$.3 with a $0^{\circ}.04$ pixel size in the 1-500 GeV energy band centered at the SIMBAD position of SNR G15.9+0.2.
 Two solid and dashed cyan circles were introduced in Figure \ref{Fig1}.  
 The green contours are from 1.4 GHz observations of NVSS \citep{Caswell1982}, and the blue contours are from X-ray observations of XMM-Newton \citep{Maggi2017}.
 A black cross represents the position (RA=274$^{\circ}$.83, decl.=-15$ ^{\circ }$.06) of 2HWC J1819-150, and its 1$\sigma$  statistical uncertainty of this position is 0$^{\circ}$.16, which is represented by a solid black circle  \citep{Abeysekara2017}. The green circle indicates the extent of the GeV emission (2D Gaussian template) in this work.
 }
 \label{Fig2}
\end{figure}

To further confirm that the region of SNR G15.9+0.2 does not have other significant $\gamma$-ray residual radiations within the $2^{\circ}.6 \times 2^{\circ}.6$ TS map, we also deducted the emission from the region of SNR G15.9+0.2. 
Using {\tt gtfindsrc}, we obtained the best-fit position of SNR G15.9+0.2 to be (R.A., decl. = 274$^{\circ}$.74, -14$^{\circ}$.92) with a 68\% (95\%) error circle of 0$^{\circ}$.10 (0$^{\circ}$.16) by assuming a point source with a power-law spectrum at its SIMBAD location. As shown in Figure \ref{Fig2},
we found its contours of the 1.4 GHz radio and X-ray energy bands are all within the 68\% and 95\% error circles of the best position of SNR G15.9+0.2.  Moreover, the region with a 1$\sigma$ statistical uncertainty from the location of the TeV source 2HWC J1819-150 (R.A., decl. = 274$^{\circ}$.83, -15$^{\circ}$.06; from \citet{Abeysekara2017}  well overlaps with all the  regions from three different energy bands.  
This indicates that these four sources from the radio to TeV bands are likely to be counterparts of SNR G15.9+0.2.

Using the uniform disk and two-dimensional (2D) Gaussian templates, we tested the different values of the radius and $\sigma$, which range from 0$^{\circ}$.1 to 1$^{\circ}$.5 with an increment of 0$^{\circ}$.05, to obtain the probable $\gamma$-ray spatial distribution of the new source. Then, we calculated the value of the TS$_{\rm ext}$, using the formula of 2log($L_{\rm ext}$/$L_{\rm ps}$) from \citet{Lande2012}, where $L_{\rm ps}$ and $L_{\rm ext}$ represent the maximum log-likelihood values for the point source template and extended templates, respectively.  
We found that the 2D Gaussian template has a significant improvement with the highest value of TS$_{\rm ext}$ and a $\sigma$ of 0$^{\circ}$.45 in the analysis. Therefore, we adopted the 2D Gaussian spatial template with a $\sigma$ of 0$^{\circ}$.45 as the best spatial template to analyze the new $\gamma$-ray source in all subsequent analyses. 
The best-fit results with the highest TS values from the different templates were presented in Table \ref{Table1}.

\begin{table}
\caption{Spatial Distribution Analysis for SNR G15.9+0.2 with Different Spatial Models in The 1-500 GeV Energy Band}
\renewcommand\arraystretch{1.}
\begin{tabular}{lcccccccc}
    \hline\noalign{\smallskip}
    Spatial Model & Radius ($\sigma$) & Spectral Index  & Photon Flux & TS
    Value & TS$_{\rm ext}$ & Degrees of Freedom  \\
                  &     degree     &      &    $\rm 10^{-9}  ph$ $\rm cm^{-2} s^{-1}$  &  &     & \\
                  
  \hline\noalign{\smallskip}
   Point source    & -             & 2.94$\pm$0.51 & 0.39$\pm$0.10 & 29.61   & - & 4 \\
   2D Gaussian        &  0$^{\circ}$.45  & 2.94$\pm$0.25 & 4.09$\pm$0.57 & 55.67 & 26.06 & 5 \\
  
   uniform disk    &  0$^{\circ}$.75 & 2.96$\pm$0.23 & 3.62$\pm$0.53 & 50.93 &  21.32 & 5\\
  \noalign{\smallskip}\hline
\end{tabular}
    \label{Table1}
\end{table}

\subsection{\rm Spectral Analysis}

\begin{figure*}
  \centering\underline{}
  \includegraphics[scale=0.6,angle=0]{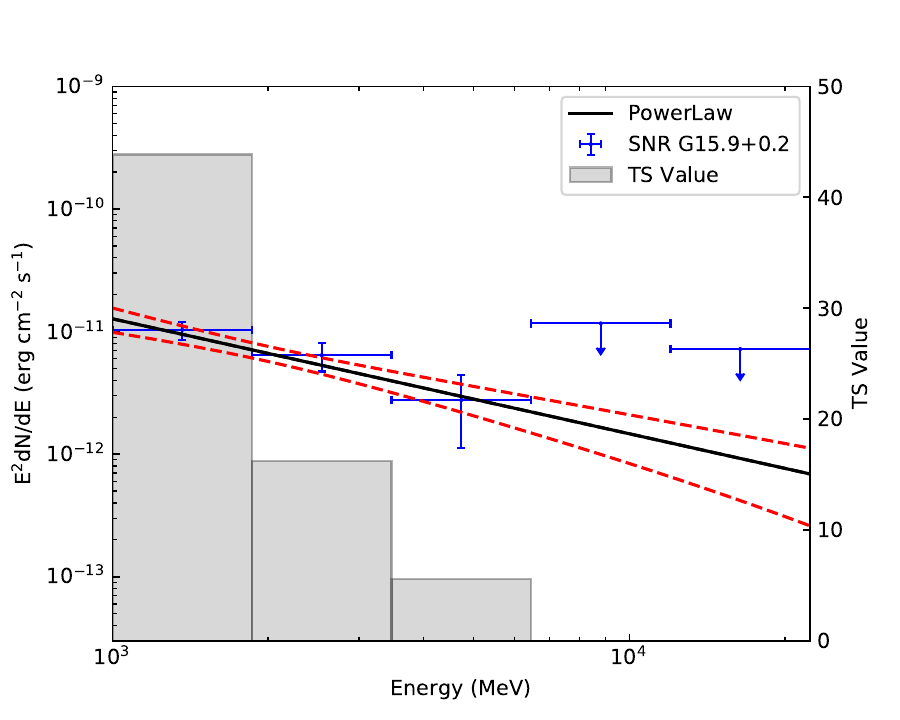}
 \flushleft
\caption{\normalsize The SED of SNR G15.9+0.2 from the 1 GeV to 500 GeV energy band. Blue points indicate the result of the Fermi-LAT observation with the 1$\sigma$ statistical uncertainties. 
 The black solid line represents the result of the global fit with a power-law spectral model, and the two red dashed lines represent the 1$\sigma$ statistical uncertainties of the global fit.
The gray shaded areas represent the TS value of each energy bin. For the energy bins with TS values $<$4, the upper limits with a 95\% confidence level are given.
}
\label{Fig3}
\end{figure*}

In this analysis, we generated the spectral energy distribution (SED) in the 1-500 GeV energy band using a power-law spectrum model with the formula, $dN/dE = N_{\rm 0}E^{-\Gamma}$,  and the 2D Gaussian template for the new $\gamma$-ray source. 
The result of the global fit is $\Gamma$= 2.94$\pm$0.25; 
its photon flux is (4.09$\pm$0.57) $\times$ $\rm 10^{-9}  ph$ $\rm cm^{-2}$ $\rm s^{-1}$. 
The SED was divided into 10 equal logarithmic bins.
Each energy bin was fitted using the binned likelihood analysis method. 
 For the energy bins with the TS value $<$ 4, we calculated their upper limits with a 95\% confidence level.  Considering the subsequent energy bins with large statistical errors and the TS values $<$ 4, here we selected to provide two upper limits for the SED, as shown in Figure \ref{Fig3}.
 
\subsection{Variability Analysis} \label{sec:data-results}

\begin{figure*}
\centering
 \includegraphics[scale=0.62,angle=0]{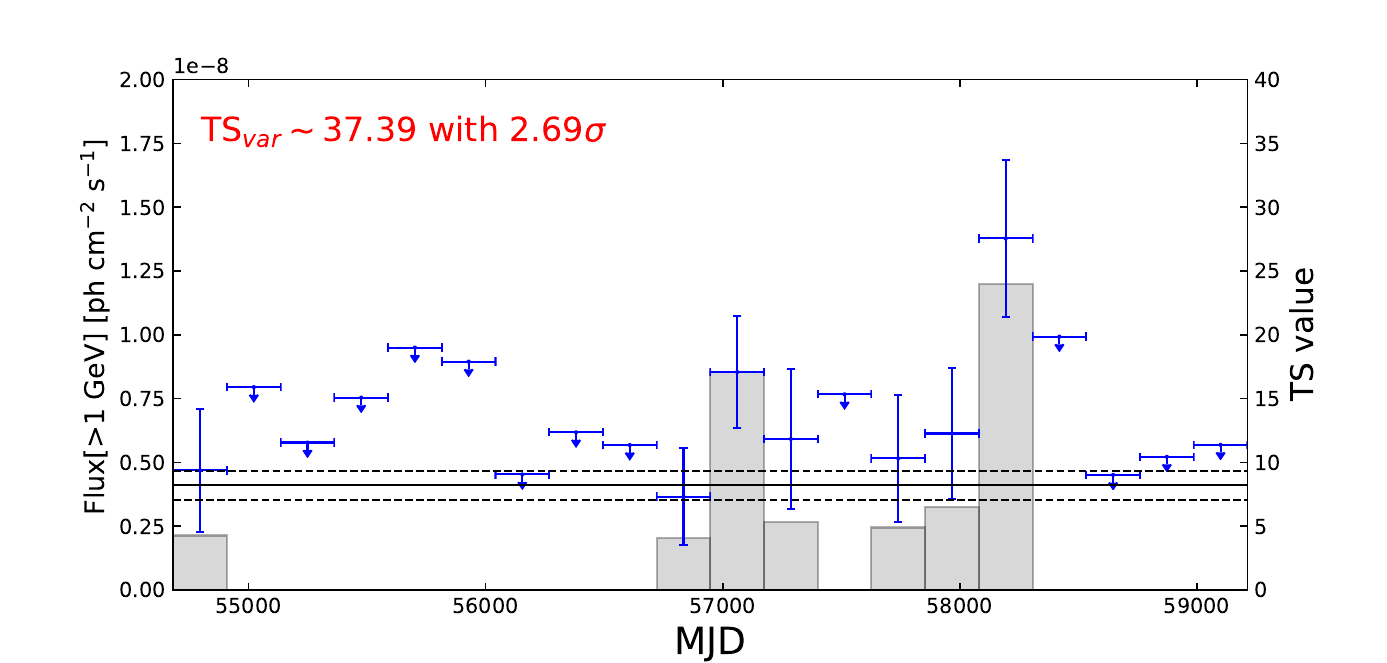} %
 \caption{\normalsize LC of SNR G15.9+0.2 with 20 time bins in the 1-500 GeV band. For the time bins  with TS values $<$ 4, the upper limits with a 95\% confidence level are given. The gray shaded areas   show the TS value of each time bin. The black solid line and the two black dashed lines are used to show the average photon flux from the maximum likelihood fit and  its 1$\sigma$ statistical  uncertainties, respectively.
}
 \label{Fig4}
\end{figure*}

To check the variability of the photon flux over 12.41 years for the new $\gamma$-ray source, we generated a light curve (LC) with 20 time bins in the 1-500 GeV energy band, as can be seen from Figure \ref{Fig4}.  
Calculating the variability index TS$_{\rm var}$ defined by \citet{Nolan2012}, we acquired $\rm TS_{var}=$ 37.39 with a 2.69$\sigma$ variability significance level\footnote{Here, the variability significance level is calculated by the SciPy package \citep{Virtanen2020}.}. 
 The result implies that the new $\gamma$-ray source exhibits a hint of weak variation\footnote{$\rm TS_{var}\geq$ 36.19 was used to identify variable sources at a 99\% confidence level for the LC of 20 time bins \citep{Xiang2021}.}.

\section{DISCUSSION} \label{sec:data-results}   

By analyzing the above TS maps of SNR G15.9+0.2, we found that the region of SNR G15.9+0.2 has the significant GeV $\gamma$-ray radiation with a 2D Gaussian spatial distribution and a significance level of 6.47$\sigma$.  Its photon flux is (4.09$\pm$0.57) $\times$ 10$^{-9}$ ph cm$^{-2}$ s$^{-1}$ with $\Gamma$= 2.94$\pm$0.25. 
We observed that almost all the radio and X-ray contours from SNR G15.9+0.2 are all within the 2$\sigma$ error circle of the best-fit location. In addition, the 1$\sigma$ error circle also contains some contours from the radio and X-ray energy bands. These results  suggest that the position of the new $\gamma$-ray source well coincides with SNR G15.9+0.2. 


Next, we analyzed the variability of approximately 12.4 years of the LC and found that the LC exits a weak variability with a significance level of 2.69$\sigma$. 
Thus far, the variability from LCs of SNRs in the Milky Way likely exists, such as iPTF14hls \citep{Yuan2018}, Supernova 2004dj \citep{Xi2020}, the Crab Nebula \citep{Arakawa2020}.
In addition, we investigated the 4FGL and found three certified SNRs with TS$_{\rm var}$ $>$ 18.48\footnote{The value of $\rm TS_{\rm var}>$ 18.48 over 12 intervals indicates that the source more than 99\% chance is a variable source. Please refer https://heasarc.gsfc.nasa.gov/W3Browse/fermi/fermilpsc.html}, including W 51C, W 44, and IC 443, which implies the variability from the LC of SNR G15.9+0.2 is likely. 
We considered that the most significant GeV radiations of the region are concentrated in the 2D Gaussian region with  0$^{\circ}$.45 radius, as displayed in Figure \ref{Fig2}, and the LC of the region has weak variability. 
Then, we used SIMBAD \footnote{simbad.u-strasbg.fr/simbad/} and Aladin\footnote{https://aladin.u-strasbg.fr/aladin.gml} to investigate whether the 2D Gaussian extended region has certified active galactic nuclei (AGN). However, we did not find likely one; therefore,  we suggest that the new GeV $\gamma$-ray emission is more likely to be from SNR G15.9+0.2.



We considered leptonic and hadronic scenarios to explain the GeV SED of SNR G15.9+0.2 from this work using a one-zone model from {\tt NAIMA} \citep[][and references therein]{Zabalza2015}. Here we assumed leptonic and hadronic particle distributions satisfy the following two particle distributions:

 (1) A power law model (PL):
\begin{equation}
 N(E) = N_{0}\left (\frac{E}{E_{0}}\right )^{-\alpha},
\end{equation}

 (2) A power law with an exponential cutoff model (ECPL):
\begin{equation}
N(E) = N_{0}\left (\frac{E}{E_{0}}\right )^{-\alpha}exp\left (-\frac{E}{E_{\rm cutoff}}\right),
\end{equation}
where $ E_{\rm 0}$ is set to 10 GeV, $N_{\rm 0}$ represents the amplitude, $E$ is the particle energy, ${\alpha}$ is the spectral index, $E_{\rm cutoff}$ represents the break energy  \citep{Aharonian2006,Ambrogi2019,Xin2019,Xiang2021}. 
In the fit, we used the Bayesian information criterion (BIC) to determine the goodness of fit of the two models \citep{Schwarz1978, Ambrogi2019}. The related formula of BIC is  ${\rm log}(n)k - 2 {\rm log}(L)$, where \textit{n} represents the number of the observed data, \textit{k} represents the number of parameters of the model, and $L$ is the  maximum likelihood value.
We assumed that the radiation fields of the leptonic and hadronic models come from the extended region of the SNR.

For the leptonic scenario, the presence of very-high-energy (VHE) electrons are confirmed in SNRs \citep[e.g.,][]{Tanimori1998}, and the GeV and TeV emissions of SNRs are also observed \citep[e.g.,][]{Ackermann2017,Abdalla2018a}.  
Subsequently, the inverse Compton scattering from leptons was widely used to explain the SEDs of high-energy bands from SNRs \citep[e.g.,][]{Tang2013,Condon2017,Zeng2017,Zeng2019,Zeng2021}. Therefore, the leptonic model, as a frequently-used model, is considered in the analysis. 
Additionally, the detection of the characteristic pion-decay signature in IC 443 and W44 confirms that cosmic-ray protons can be  accelerated in SNRs \citep{Ackermann2013}. 
The proportion of proton composition of the observed CR spectrum on Earth is 99\%, suggesting that the hadronic contribution for the $\gamma$-ray emission from SNRs cannot be ignored \citep{Liu2015}. Recent studies have shown that the multi-band SED of certain SNRs can be better explained when considering the contribution of hadrons than a pure leptonic scenario. E.g., Puppis A \citep{Xin2017}, SNR G106.3+2.7 \citep{Xin2019,Yang2021}, Kepler's SNR \citep{Xiang2021}.  
Consequently, we also considered the hadronic origin here.

\begin{figure*}
\centering
 \includegraphics[width=\textwidth, angle=0,width=90mm,height=90mm]{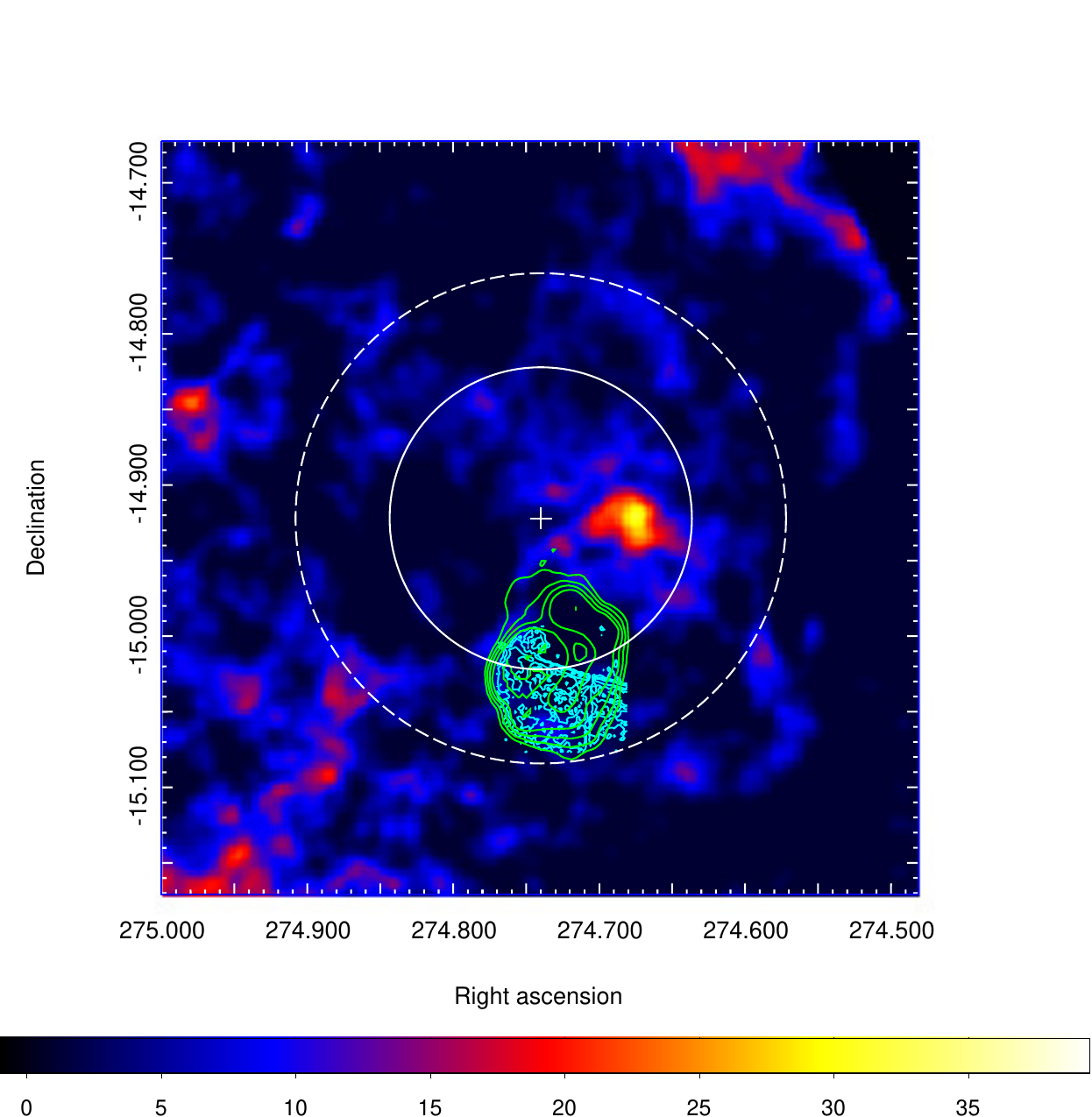} %
 \caption{\normalsize The 0$^{\circ}$.5$\times$0$^{\circ}$.5 velocity-integrated brightness temperature map, smoothed with a Gaussian kernel of 0$^{\circ}$.3, from the CO Milky Way Survey \citep{Dempsey2013}. The color bar represents CO intensity with the units of K km s$^{-1}$. 
The 68\% and 95\% error circles of the best-fit position of this 
SNR are marked by using solid and dashed white circles in these TS maps, respectively. The green contours are from the observation of NVSS. The cyan contours are from the observation of XMM-Newton.}
 \label{Fig6}
\end{figure*}

The interaction between high-energy particles of SNR and the surrounding molecular cloud can serve as the important evidence for the hadronic origin \citep{Wootten1977,Denoyer1979,Tatematsu1990,Green1997,Reach1999,Zhou2009,Kilpatrick2014}.
 Here, we firstly investigated OH maser emission at 1720 MHz around SNR G15.9+0.2 from \citet{Green1997}. However, we did not find the significant OH maser emission around SNR G15.9+0.2; thus, there is no convincing evidence to verify the interactions of SNR G15.9+0.2 with ``OH'' molecular clouds. 
 In addition, \citet{Tian2019} proposed that the region of SNR G15.9+0.2 is likely to produce high-energy emissions, owing to the likely interactions between SNR G15.9+0.2 and the surrounding CO molecular clouds. 
Subsequently, we quickly investigated the past CO observations from  \citet{Dempsey2013}; we found dense CO molecular clouds in the 1$\sigma$ error circle of the best-fit position of SNR G15.9+0.2, as shown in Figure \ref{Fig6},  which implies that this high-energy $\gamma$-ray radiation may originate from interactions between the high-energy particles of SNR G15.9+0.2 and the target particles of the nearby CO molecular clouds.

 However, no broad CO emission lines were observed from the region in their works.  Therefore, the interaction between SNR and surrounding CO molecules is uncertain and needs further observations to verify in the future. 


Owing to the uncertainty of the interaction with the surrounding CO molecules, we assumed that the GeV radiation of the SNR comes from the interaction between the high-energy particles inside the forward shock and the surrounding interstellar medium of the SNR, and its related gas density $n_{\rm gas}$= 0.7 cm$^{-3}$ \citep{Reynolds2006}.  For the distance of SNR G15.9+0.2, we took 8.5 kpc from \citet{Reynolds2006} for fitting the SED.  
Here, {\tt PYTHIA 8}, which involves the cross-section of proton-proton interactions and pion production from \citet{Kafexhiu2014}, was selected to perform the analysis. 

\begin{figure*}
\begin{center}
  \includegraphics[scale=0.5,angle=0]{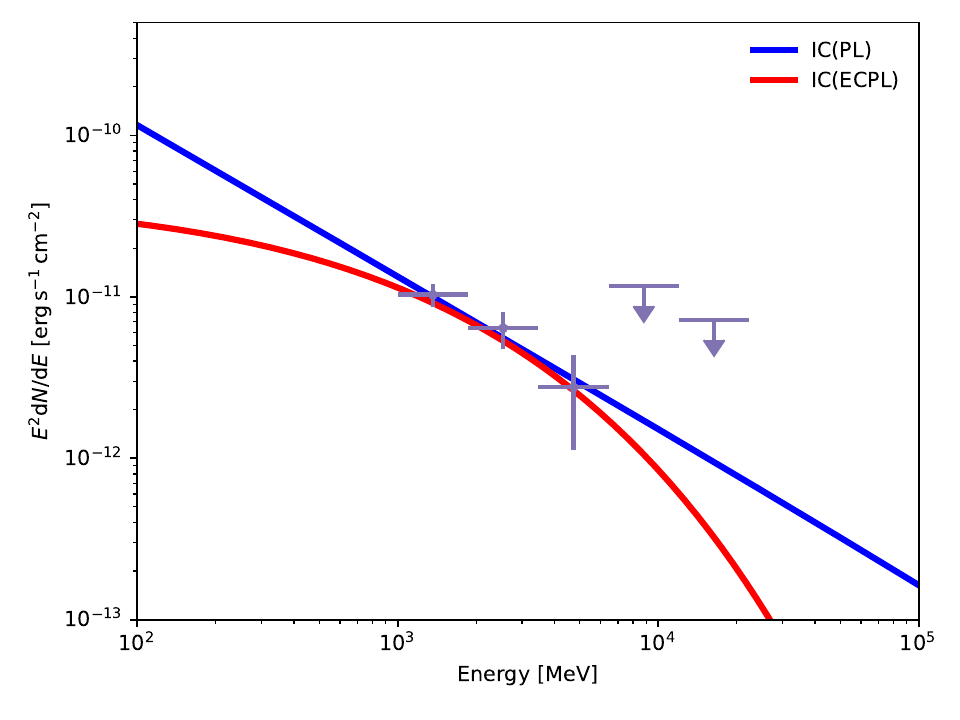}
  \includegraphics[scale=0.5,angle=0]{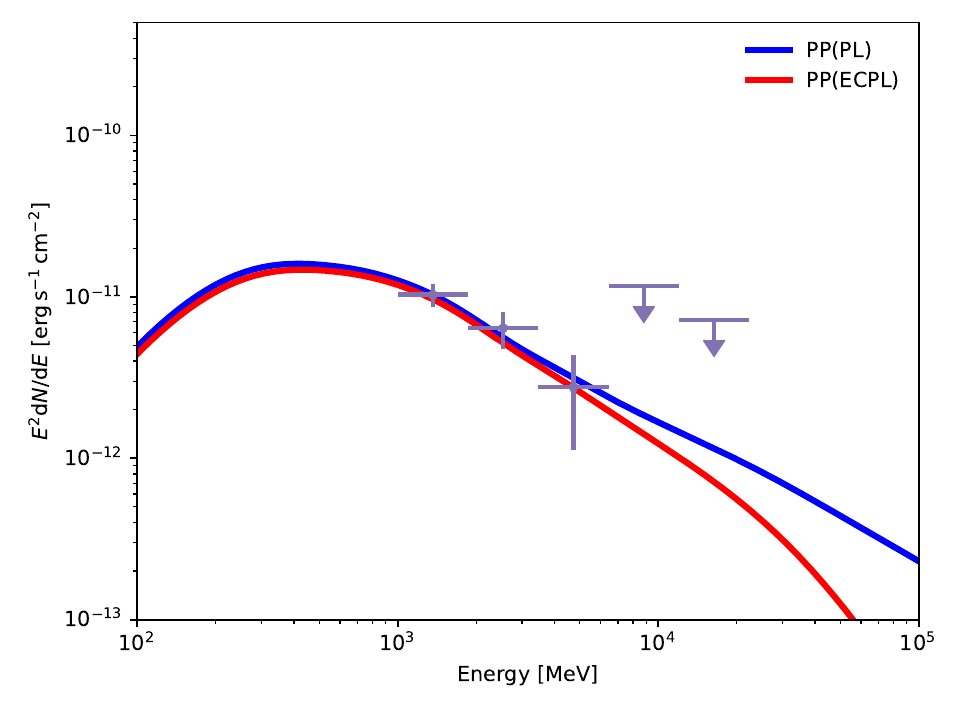}
 \flushleft
\caption{\normalsize 
For the above two panels, the blue and red solid lines represent the best-fit results of PL and ECPL, respectively. Two upper limits are included in the fitting process. 
Left panel: the leptonic scenario that is dominated by the inverse Compton scattering. Right panel: the hadronic scenario that is dominated by the decay of neutral pions from the process of proton-proton interactions. 
}
\label{Fig7}
\end{center}
\end{figure*}

The best-fit results of the two radiation models are shown in the two panels of Figure \ref{Fig7} and Table \ref{Table2},  
we found that leptonic and hadronic models with particle distributions  of PL and ECPL can explain the SED with a reduced $\chi^{2}$ value of  approximately 0.  
Moreover, we found that BIC values of PL and ECPL are close, with the values of approximately 3 and 5, respectively. Therefore, the goodness of fit of PL and ECPL cannot be distinguished thus far.
The actual particle distribution of the SNR needs more high-energy observation data to infer in the future (e.g., continuous Fermi-LAT observation above 7 GeV). Furthermore, we found that both PL and ECPL models have soft indexes for leptonic and hadronic scenarios. Since the timescale of particle acceleration in SNR is    approximately hundreds to thousands of years \cite {Yuan2018}, considering the current age of the SNR is approximately 1000-3000 years \citep{Reynolds2006},  this may prevent the particles in the SNR from being accelerated to a very high energy level, resulting in a rapid truncation above 1 GeV and presenting a soft spectral feature.

\begin{table*}
\caption{The Best-fit Parameters of Leptonic and Hadronic Models}
\begin{center}
\renewcommand\arraystretch{1.5}
\begin{tabular}{cccccccc}

    \hline\noalign{\smallskip}
  Model Name    & Particle Distribution  & $N_{\rm 0}$   & $\alpha$               & $E_{\rm cutoff}$        & -log$(L)$ &   BIC & $\chi^{2}/N_{dof}$   \\
                        &  &  & & GeV  &  &       &  \\
  \hline\noalign{\smallskip}
  \multirow{2}{*}{Leptonic model} & PL  & $4.81^{+1.08}_{-1.25}$ $\times 10^{45}$ & $4.87^{+0.07}_{-0.09}$ & --- & -0.11 & 3.44 & $\frac{0.11*2}{5-2}=0.07$   \\
  
     & ECPL  &  $2.44^{+0.55}_{-0.66}$ $\times 10^{43}$ &   $2.94^{+0.20}_{-0.23}$  & $243.46^{+89.64}_{-68.42}$   & -0.03 &  4.89 &   $\frac{0.03*2}{5-3}=0.03$ \\ 
  \noalign{\smallskip}\hline 
    \multirow{2}{*}{Hadronic model} & PL   & $ 5.74^{+0.52}_{-0.60}$ $\times 10^{42}$ & $2.99^{+0.22}_{-0.20}$    & --- & -0.13 & 3.48 &  $\frac{0.13*2}{5-2}=0.09$   \\
     & ECPL   & $6.17^{+0.93}_{-1.00}$ $\times 10^{42}$  & $ 2.85^{+0.44}_{-0.40}$  & $201.81^{+60.81}_{-68.14}$  & -0.10 & 5.03 &   $\frac{0.10*2}{5-3}=0.10$ \\  
    \noalign{\smallskip}\hline  
\end{tabular}
\end{center}
\label{Table2}  
 \flushleft \normalsize
\end{table*}



\subsection{\rm Summary}
   \begin{enumerate}
      \item 
By analyzing approximately 12 years of Fermi-LAT Pass 8 data in the 1-500 GeV energy band, a new GeV source with 6.47$\sigma$ is found from the region of SNR G15.9+0.2; its photon flux is (4.09$\pm$0.57)$\times$ $\rm 10^{-9}  ph$ $\rm cm^{-2} s^{-1}$ with a soft spectral index of 2.94$\pm$0.25.

      \item Its spatial distribution can be described by a 2D Gaussian spatial model with $\sigma=0^{\circ}.45$.
  
      \item The GeV spatial position of SNR G15.9+0.2 is in good agreement with those of the radio, X-ray, and TeV bands. This result suggests that the new GeV source is likely to be a counterpart of SNR G15.9+0.2.      
      
      \item Its LC presents a weak varibility, which is likely for the currently observed SNRs in Milky Way.      
      
      \item Its SED can be explained by considering the leptonic and hadronic scenarios with PL and ECPL particle distribution models. For the hadronic origin, past observations show that there are dense CO molecular clouds around it, but such proton-proton interactions need to be further demonstrated in the future.

   \end{enumerate}

\section{Acknowledgments} 
We sincerely thank the referee for his/her invaluable 
comments and appreciate the support for this work from
the National Key R\&D Program of China under Grant No. 2018YFA0404204, the National Natural Science Foundation of China (NSFC U1931113, U1738211, 11303012), the Foundations of Yunnan Province (2018IC059, 2018FY001(-003), 2018FB011), the Scientific Research Fund of Yunnan Education Department (2020Y0039).

\end{document}